\documentstyle [12pt,aaspp4]{article}
\def\gs{\mathrel{\raise1.16pt\hbox{$>$}\kern-7.0pt
\lower3.06pt\hbox{{$\scriptstyle \sim$}}}}
\def\ls{\mathrel{\raise1.16pt\hbox{$<$}\kern-7.0pt
\lower3.06pt\hbox{{$\scriptstyle \sim$}}}}
\def\mpc{\,h^{-1}{\rm Mpc}}
\def\pppm{\rm P^3M}

\def\kms{\,{\rm {km\, s^{-1}}}}

\def\bx{{\vec {x}}}
\def\br{{\vec {r}}}
\def\bv{{\vec {v}}}

\begin{document}

\title{The Pairwise Peculiar Velocity Dispersion of Galaxies: Effects
  of the Infall}

\author {Y.P. Jing$^{1}$, G. B\"orner$^{1,2}$}
\affil{$ ^1$Research Center for the Early Universe,
School of Science, University of Tokyo, Bunkyo-ku, Tokyo 113, Japan}
\affil {$ ^2$Max-Planck-Institut f\"ur Astrophysik,
Karl-Schwarzschild-Strasse 1, 85748 Garching, Germany}
\affil {e-mail: jing@utaphp2.phys.s.u-tokyo.ac.jp, ~ grb@mpa-garching.mpg.de}
\received{---------------}
\accepted{---------------}
\begin{abstract}

  We study the reliability of the reconstruction method which uses a
  modelling of the redshift distortions of the two-point correlation
  function to estimate the pairwise peculiar velocity dispersion of
  galaxies. In particular, the dependence of this quantity on different
  models for the infall velocity is examined for the Las Campanas
  Redshift Survey. We make extensive use of numerical simulations and
  of mock catalogs derived from them to discuss the effect of a
  self-similar infall model, of zero infall, and of the real infall
  taken from the simulation. The implications for two recent
  discrepant determinations of the pairwise velocity dispersion for
  this survey are discussed.

\end{abstract}

\section{Introduction}

The Pairwise peculiar Velocity Dispersion (PVD) of galaxies is in
principle a well-defined statistical quantity which can give
interesting information on the cosmic matter distribution in addition
to the two-point correlation function. The peculiar velocities of
galaxies are determined by the action of the local gravitational
fields, and thus they directly mirror the gravitational potentials
caused by dark and luminous matter. The PVD is measured by modelling
the distortions in the observed redshift-space correlation function
$\xi_z(r_p, \pi)$ which is in general not just a function of the
distance $ s^2 = r_p^2 + \pi^2 $, but depends anisotropically on the
separations of a galaxy pair perpendicular to ($r_p$) and along
($\pi$) the line of sight. This is the information that can be
obtained from a redshift survey. The basic step in modelling is to
write $\xi_z(r_p, \pi)$ as a folding integral of the real-space
correlation function $\xi(r)$ and the distribution function $
f(v_{12})$ of the relative velocity $v_{12}$ of galaxy pairs along the
line of sight
\begin{equation}\label{xizmodel}
1+\xi_z(r_p, \pi)=
\int
f(v_{12})\left[1+\xi(\sqrt{r_p^2+(\pi-v_{12}/H_0)^2})
\right]dv_{12}, 
\end{equation} 
(see, e.g., Davis \& Peebles \cite{davis}).  The real space correlation
function $ \xi(r)$ must be estimated from the redshift catalog through
the relation
\begin{equation}\label{wrp}
w(r_p)=\int_0^\infty \xi_z(r_p, \pi) d\pi=
\int_0^\infty \xi(\sqrt{r_p^2+y^2}) dy
\end{equation}
where $ w(r_p) $ is the so-called `projected' two-point correlation
function. In most previous works a power-law form is assumed for
$\xi(r)$.  Based on observational (Davis \& Peebles \cite{davis}; Fisher et al
\cite{fisher}) and theoretical considerations (Diaferio \& Geller \cite{diaferio}; Sheth
\cite{sheth}; Seto \& Yokoyama \cite{seto}) an exponential form is usually adopted
for $f(v_{12})$:
\begin{equation}\label{fv12}
f(v_{12})={1\over \sqrt{2}\sigma_{12}} \exp \left(-{\sqrt{2}\over
\sigma_{12}} \left| v_{12}-\overline{v_{12}}\right| \right), 
\end{equation}
where $\overline{v_{12}}$ is the mean and $\sigma_{12}$ is the
dispersion of the 1-D pairwise peculiar velocities along the line of
sight. It is worth pointing out that every step in the above modelling
[eqs. (\ref{xizmodel}-\ref{fv12})] is only an approximation and the
infall $\overline{v_{12}}$ is unknown. As demonstrated by Jing, Mo, \&
B\"orner (\cite{jing98}; JMB) (see
also below) with mock samples of the Las Campanas Redshift Survey
(LCRS; Shectman et al. \cite{shectman}), however, the above procedure
can give an accurate estimate of $\sigma_{12}$ (within 20\% accuracy)
if the infall is known.

The distribution function $f(v_{12})$ is determined by its first and
second moment: the infall ($\overline{v_{12}}$) and the dispersion
$\sigma_{12}(r)$. Both distort the two-point correlation function but in
opposite ways. The infall velocity must also be modelled in some
detail to allow a precise measurement of the dispersion
$\sigma_{12}(r)$. The situation seems somewhat complex:
$\overline{v_{12}}$ in the real Universe is not known at the present.
One might think that on small scales $\overline{v_{12}}$ is
negligible, but this is true only for very small scales indeed. As has
been shown by Mo, Jing, \& B\"orner (\cite{mo}) the function
$\overline{v_{12}}$ rises quite sharply around $1\mpc$, reaching twice
the Hubble velocity just beyond $1\mpc$. Therefore it is necessary to
model the infall carefully when measuring $\sigma_{12}(r)$.

In JMB we have determined the PVD for the LCRS using this
reconstruction method. Recently an attempt to measure the PVD for the
same survey directly from a Fourier deconvolution of the anisotropies
of the redshift space two-point correlation function (Landy et al
\cite{landy}) has resulted in a much lower value for the PVD ($363\pm 44\kms$
vs $570 \pm 80\kms$). We think it is important to identify the causes
for the discrepancy, since the PVD is a very powerful test for the
theories of the structure formation. We shall show that the different
infall models used in the two studies can explain the discrepancy.
But in addition to this immediate aspect there is the
general question of how reliable these methods to measure the PVD
really are. In this paper we want to address this question.

\section{N-Body Simulation and mock samples}
The true PVD can be easily determined from the three-dimensional
velocities of particles in numerical simulations. Writing the
three-dimensional velocity difference of particle pairs at points $ \bx
$ and $\bx + \br$, i.e. at separation $\br$, as
\begin{equation}\label{v12}
\bv_{12}(\br)=\bv (\bx)-\bv (\bx+\br)
\end{equation}
the true PVD is defined as
\begin{equation}\label{sig3d}
\sigma_{12}(r)=\Bigl\langle \bigl(\bv_{12}(\br)-\langle
\bv_{12}(\br)\rangle
\bigr)^2/3\Bigr\rangle^{1/2}
\end{equation}
where $\langle\cdot\cdot\cdot\rangle$ denotes the average over all
pairs at separation $r$. In JMB we have considered three spatially
flat cosmological models. Here we make use of one simulation of the
model with density parameter $\Omega_0=0.2$, cosmological constant
$\lambda_0 = 0.8 $, shape parameter $ \Gamma = 0.2$ and normalization
$ \sigma_8 =1 $. This N-body simulation was generated with a $\pppm$
code (Jing and Fang \cite{jing94}) with $128^3$ particles. Twenty mock
samples without the fiber collisions are generated from the simulation
to mimic the LCRS.  We use these mock samples to determine the PVD in
the way outlined above, and compare the results to the true PVD
obtained from the simulations. Further details about the simulation,
the mock samples, and our statistical method were given in JMB.

\section{Dependence on Infall-models}

\begin{figure}
\epsscale{1.0} \plotone{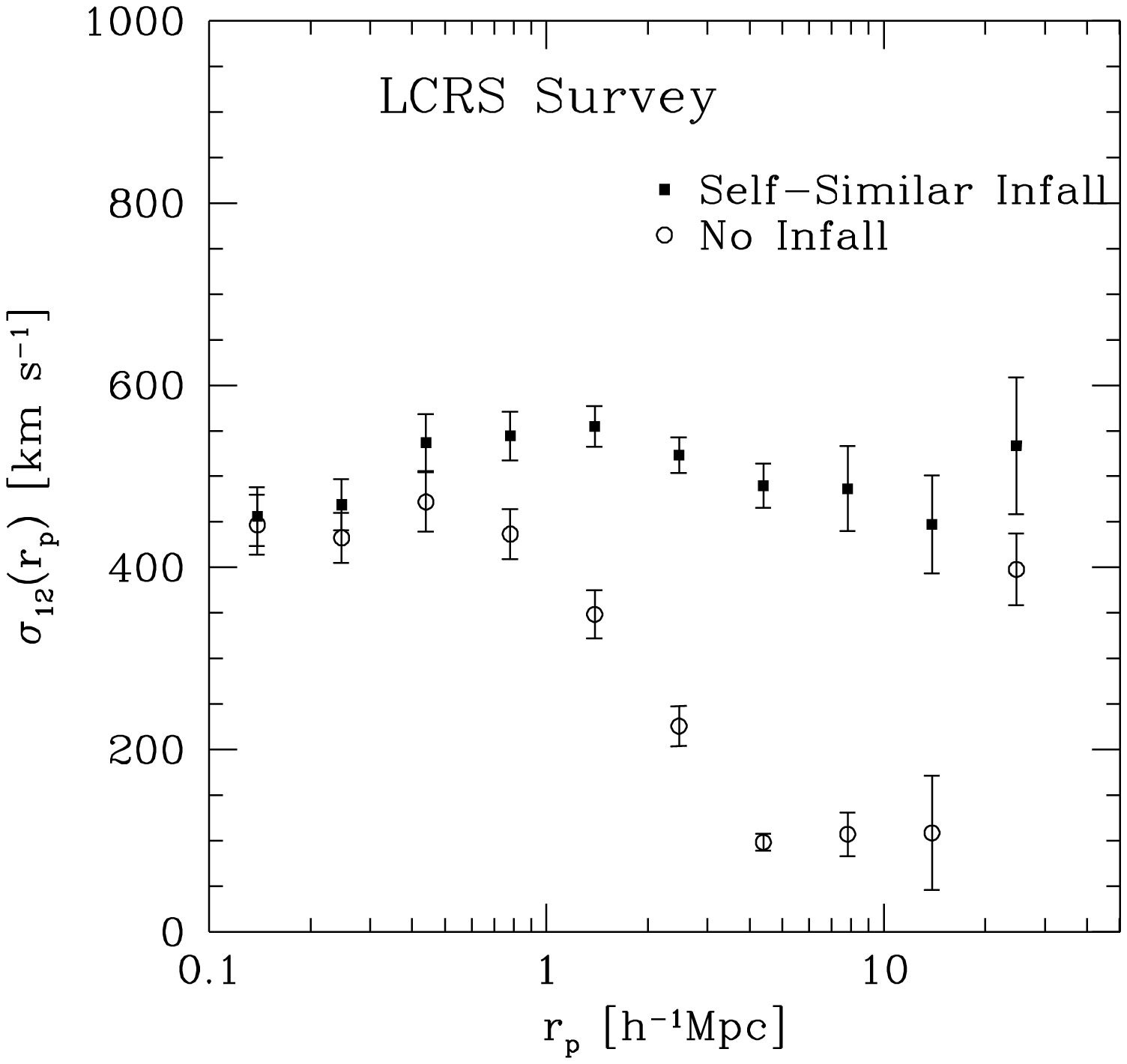}
\caption{  The PVD of
  the Las Campanas Redshift Survey: filled squares for the
  self-similar infall and circles for no infall. Error bars are
  $1\sigma$ deviations given by bootstrap resampling. }
\label{fpvdo}
\end{figure}

\begin{figure}
\epsscale{1.0} \plotone{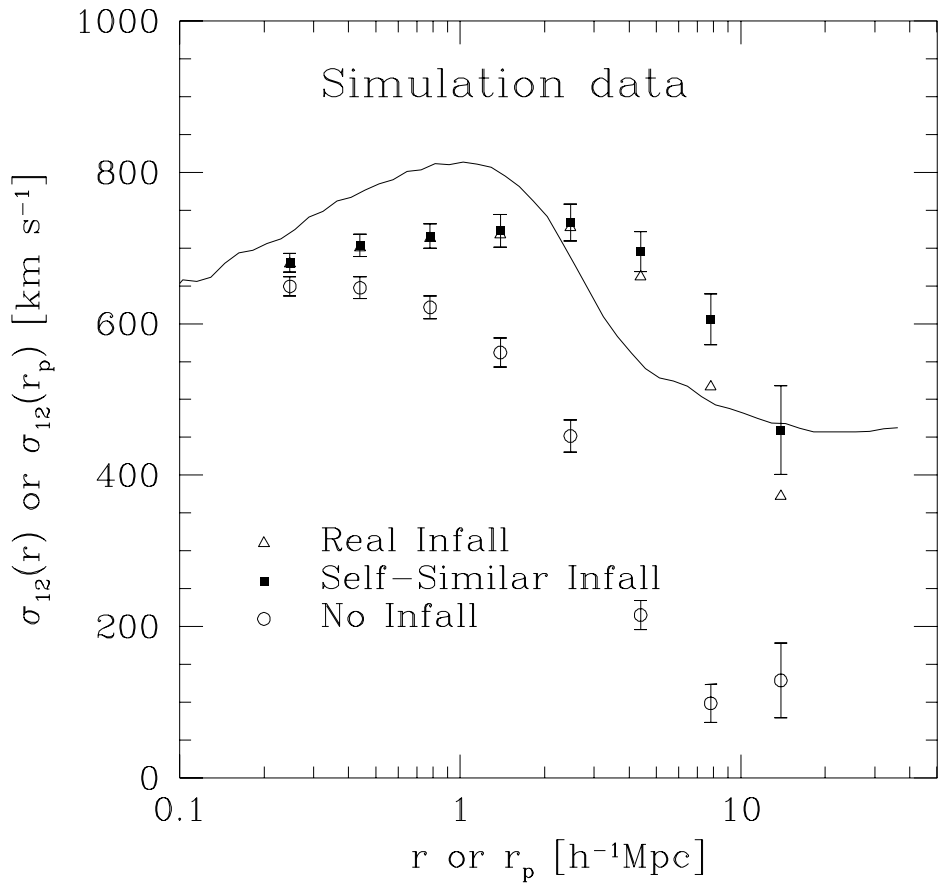}
\caption{  The PVD of 20 mock samples without fiber collisions.
  Three infall models are adopted for $\overline {v_{12}}(r)$: the
  self-similar infall model (filled squares), the zero infall model
  (circles), the infall derived directly from the simulations (open
  triangles).  The {\it true} pairwise velocity dispersion given by
  the 3-dimensional velocities in the simulations is shown as the
  solid line.  Error bars are the ($1\sigma$) standard deviations of
  the mean from the mock samples.  }
\label{fpvds}
\end{figure}

In JMB we have determined the PVD for the LCRS and found a best
estimate of
\begin{equation}\label{sigres1}
\sigma_{12}(1\mpc) = 570 \pm 80\kms
\end{equation}
at a separation of $1\mpc$. Figure~\ref{fpvdo} gives an indication of
the reliability of the result: accepting the exponential shape of the
distribution function $f(v_{12})$ as a reasonable ansatz, there
remains the mean value $v_{12}(r)$ to be modelled.  The filled squares
in Fig.~\ref{fpvdo} represent the results from the modelling process,
when a self-similar infall model is used for $\overline {v_{12}}$:
\begin{equation}\label{ssinf}
\overline{v_{12}}({\bf r})= -{y H_0 \over 1+(r/r_{\star})^2};
\hskip 1cm r_{\star}=5\mpc, 
\end{equation}
and $y$ is the radial separation in real space. This form for
$\overline{v_{12}} $ has been widely used in previous work.  It also
is a good approximation to the real infall pattern in some CDM models.
The circles in Fig.~\ref{fpvdo} denote the PVD reconstructed when the
infall is set to zero.  In Fig.~\ref{fpvds}, we have plotted the true
PVD read off directly from the simulation as the solid line. Circles
are the result from the mock catalogs with zero infall; triangles
depict the result from the modelling process when the real infall from
the simulation is used, and filled squares are again the
reconstruction for a self-similar infall.  The two different infall
models give very similar results on scales $ r_p < 10 \mpc$.  We may
conclude that the reconstruction of the PVD does not depend very
sensitively on the model for $\overline{v_{12}}$.  The PVD
reconstructed from the redshift distortions agrees qualitatively with
the true value. There are, however, differences of some significance,
even if the real infall pattern is used, which reflects the
approximate features of the modelling
[eqs. (\ref{xizmodel}-\ref{fv12})]. The model, where the infall is
completely neglected, in contrast does not even qualitatively
correspond to the true value. Although the infall velocity becomes
negligible on small scales it still has a strong influence on the PVD:
at $1\mpc$ we find a reduction from $570\kms$ to $400\kms$, if
$\overline{v_{12}}=0$, and at larger scales the no-infall model lets
the PVD go down quite rapidly.  The same behaviour of rapid drop of
the no-infall models can be seen in the simulation results
(Fig.~\ref{fpvds}). (The simulation results are higher, because we
have not yet applied the bias necessary to achieve agreement with the
observation). It is probably due to the shape of the infall velocity
(Mo et al. \cite{mo}) around $1\mpc$ with a steep rise, and a maximum
at a few $\mpc$ for all CDM models considered in Mo et
al. (\cite{mo}). The Fourier deconvolution method as applied by Landy
et al. (\cite{landy}) appears simpler, because it does not seem
to have to model the infall $\overline{v_{12}}$. In fact, however,
they use the model $\overline{v_{12}}=0$ which can, as we have seen from
our simulations, lead to an underestimate of the true value. To obtain the
true PVD therefore also the Fourier deconvolution method needs to
model the infall, and thus it meets the same difficulties as the usual
approach.  

The influence on the PVD estimate at $\sim 1\mpc$ of the infall
models we found here is in qualitative agreement with many previous
works, e.g, Davis \& Peebles (1983) and Marzke et
al. (\cite{marzke}). Quantitatively the influence may depend on the sample
used, which is particularly true for small samples, since the infall
effect depends on which regions (clusters or fields) the sample has
surveyed. Certainly the infall effect should be universal for {\it
  fair} samples, but it is not known that any observational sample
available to date could be considered ``fair'' for the infall effect.
Therefore it is necessary to quantify this effect individually for
each sample, as we have done for the LCRS here.

\section{Discussion}
The modelling of the redshift distortions of the two-point correlation
function gives a reasonable estimate ---certainly within 20\% of the
true PVD, despite the complex reconstruction method involved. The
differences are due to the fact that the form of $ f(v_{12}) $ is only
approximately an exponential, and that the PVD estimated from the
redshift distortions is some kind of average of the true PVD along the
line of sight. Since the true PVD depends on the separation of galaxy
pairs in real space, these two quantities are different by definition.
Whether we use the self-similar or the true infall model has little
effect on the results. It is very important , however, to use both the
first and second moments of the velocity distribution function in the
modelling process, since they lead to distortions of the redshift
space correlations in opposing directions. Thus, for instance setting
$\overline{v_{12}}=0$ leads to drastic changes. $ \sigma_{12} $ at
$1\mpc $ drops from 570 to about 400 km/s, and it becomes very small
even for $\gs 1 \mpc $. We suspect that this behaviour is responsible
for the result of a recent work (Landy et al \cite{landy}), where $\sigma_{12}
$ from the LCRS is estimated to be $363 \pm 44 \kms$ at scale $\sim
1\mpc$. Their reconstruction method makes use of a Fourier
deconvolution of $ \xi_z(r_p, \pi) $ , but the effects of infall are
neglected. We can reproduce their result, if we set
$\overline{v_{12}}=0$ , but from our comparison between simulations
and mock catalogues we can draw the conclusion that this approach 
may have underestimated the true PVD and that the analysis incorporating
a reasonable infall, like that of JMB, gives a much more reliable result.

Considering the fact that the two works have used rather different
methods to measure the PVD, we would stress that one should remain open
minded with respect to both values even though we have quantitatively explained the
discrepancy with the different infall models. However, the PVD
is a very important quantity in cosmology. Since the procedure of
JMB has been extensively tested with mock samples, it is important and
necessary to make a similar test to the procedure of Landy et al. 
(\cite{landy}).

\acknowledgements
  We are grateful to Yasushi Suto for helpful discussions, and for the
  hospitality extended to us at the physics department of Tokyo
  university.  G. B. thanks the Yamada foundation for support during
  his stay at RESCEU. J.Y.P. gratefully acknowledges the receipt of a
  JSPS postdoctoral fellowship. Support from SFB375 is also acknowledged.           The simulations were carried out on
  VPP/16R and VX/4R at the Astronomical Data Analysis Center of the
  National Astronomical Observatory, Japan.

\end{document}